# The Impact of Correlated Metrics on Defect Models

Jirayus Jiarpakdee, *Student Member, IEEE,* Chakkrit Tantithamthavorn, *Member, IEEE,*
and Ahmed E. Hassan, *Member, IEEE*

**Abstract**—Defect models are analytical models that are used to build empirical theories that are related to software quality. Prior studies often derive knowledge from such models using interpretation techniques, such as ANOVA Type-I. Recent work raises concerns that prior studies rarely remove correlated metrics when constructing such models. Such correlated metrics may impact the interpretation of models. Yet, the impact of correlated metrics in such models has not been investigated. In this paper, we set out to investigate the impact of correlated metrics, and the benefits and costs of removing correlated metrics on defect models. Through a case study of 15 publicly-available defect datasets, we find that (1) correlated metrics impact the ranking of the highest ranked metric for all of the 9 studied model interpretation techniques. On the other hand, removing correlated metrics (2) improves the consistency of the highest ranked metric regardless of how a model is specified for all of the studied interpretation techniques (except for ANOVA Type-I); and (3) negligibly impacts the performance and stability of defect models. Thus, researchers must (1) mitigate (e.g., remove) correlated metrics prior to constructing a defect model; and (2) avoid using ANOVA Type-I even if all correlated metrics are removed.

# 1 INTRODUCTION

Defect models are constructed using historical project data to estimate the risk of having future defects in modules. However, another key usage of defect models is analytical in nature. Such analytical models are often used to explore the impact of various phenomena on software quality and to build empirical theories that are related to software quality through the use of such models to test various hypotheses. Plenty of prior studies investigate the impact of many phenomena on code quality using software metrics, for example, code size [1], code complexity [38, 54, 81], change complexity [37, 46, 64, 66, 81, 96], antipatterns [44], developer activity [81], developer experience [68], developer expertise [7], developer and reviewer knowledge [87], design [4, 13, 14, 17, 21], reviewer participation [56, 88], code smells [43], and mutation testing [9].

To perform such studies, there are five common steps: (1) formulating of hypotheses that pertain to the phenomena that one wishes to study; (2) designing appropriate metrics to operationalize the intention behind the phenomena under study; (3) defining a model specification (e.g., the ordering of metrics) to be used when constructing an analytical model; (4) constructing an analytical model using, for example, regression models [7, 64, 87, 88, 95] or random forest models [30, 42, 61, 70]; and (5) examining the ranking of metrics using a model interpretation technique (e.g., ANOVA Type-I, one of the most commonly-used interpretation techniques since it is the default built-in function for

- *J. Jiarpakdee and C. Tantithamthavorn are with the School of Computer Science, University of Adelaide, Australia.*
  *E-mail: {jirayus.jiarpakdee, chakkrit.tantithamthavorn}@adelaide.edu.au.*
- *A. E. Hassan is with the School of Computing, Queen's University, Canada.*
  *E-mail: ahmed@cs.queensu.ca.*

logistic regression (`glm`) models in R) in order to test the hypotheses.

For example, to study whether complex code increases project risk, one might use the number of reported bugs (`bugs`) to capture risk, and the McCabe's cyclomatic complexity (`CC`) to capture code complexity, while controlling for code size (`size`). We note that one needs to use control metrics to ensure that findings are not due to confounding factors (e.g., large modules are more likely to have more bugs). Then, one must construct an analytical model with a model specification of `bugs~size+CC`. One would then use an interpretation technique (e.g. ANOVA Type-I) to determine the ranking of metrics (i.e., which metrics have a strong relationship with bugs).

Metrics of prior studies are often correlated [29, 39, 50, 51, 84, 93]. For example, Landman *et al.* [50, 51], Herraiz *et al.* [39], and Gil *et al.* [29] point out that code complexity (`CC`) is often correlated with code size (`size`). Zhang *et al.* [93] point out that many metric aggregation schemes (e.g., averaging or summing of McCabe's cyclomatic complexity values at the function level to derive file-level metrics) often produce correlated metrics.

Recent studies raise concerns that correlated metrics may impact the interpretation of defect models [84, 93]. Indeed, our motivating analyses (Section 3) show that simply rearranging the ordering of correlated metrics in the model specification (e.g., from `bugs~size+CC` to `bugs~CC+size`) would lead to different ranking of metrics—the importance scores are sensitive to the ordering of correlated metrics in a model specification. Thus, if one wants to show that code complexity is strongly associated with risk in a project, one simply needs to put code complexity (`CC`) as the first metric in their models (i.e., `bugs~CC+size`), even though a more careful analysis would show that `CC` is not associated with `bugs` at all. The sensitivity of the model specification when correlated metrics are included in a model is a critical problem, since the contribution of many prior studies can be altered by simply re-ordering metrics in the model specification if correlated metrics are not properly mitigated. Unfortunately, a literature survey of Shihab [77] shows that as much as 63% of defect studies that are published during 2000-2011 do not mitigate (e.g., removing) correlated metrics prior to constructing defect models.

In this paper, we set out to investigate (1) the impact of correlated metrics on the interpretation of defect models; (2) the benefits of removing correlated metrics on the interpretation of defect models; and (3) the costs of removing correlated metrics on the performance and stability of defect models. In order to detect and remove correlated metrics, we apply the variable clustering (VarClus) and the variance inflation factor (VIF) techniques. We construct logistic regression and random forest models using mitigated (i.e., no correlated metrics) and non-mitigated datasets (i.e., not treated). Finally, we apply 9 model interpretation techniques, i.e., ANOVA Type-I, 4 test statistics of ANOVA Type-II (i.e., Wald, Likelihood Ratio, F, and Chi-square), scaled and non-scaled Gini Importance, and scaled and non-scaled Permutation Importance. We then compare the performance and interpretation of defect models that are constructed using mitigated and non-mitigated datasets. Through a case study of 15 publicly-available defect datasets of systems that span both proprietary and open source domains, we address the following four research questions:

**(RQ1) How do correlated metrics impact the interpretation of defect models?**
Irrespective of the built-in interpretation techniques for logistic regression and random forest, correlated metrics introduce inconsistency to the ranking of the highest ranked metric, highlighting the risks of not mitigating correlated metrics before constructing models.

**(RQ2) After removing all correlated metrics, how consistent is the interpretation of defect models among different model specifications?**
After removing all correlated metrics, the highest ranked metric according to Type-II, Gini Importance, and Permutation Importance is consistent. However, the highest ranked metric according to Type-I is inconsistent (as the ranking of a metric is impacted by its order in the model specification when analyzed using Type-I, the default analysis for the `glm` model in R, which is commonly used in prior studies).

**(RQ3) After removing all correlated metrics, how consistent is the interpretation of defect models among the studied interpretation techniques?**
After removing all correlated metrics, we find that at least one metric in the top-3 ranked metrics of the studied model interpretation techniques is consistent for 87%-100% of the studied datasets, highlighting the benefits of removing all correlated metrics on the interpretation of defect models, i.e., the conclusions of studies that rely on one interpretation technique may not pose a threat after mitigating (e.g., removing) correlated metrics.

**(RQ4) Does removing all correlated metrics impact the performance and stability of defect models?**
Removing all correlated metrics decreases the AUC, F-measure, and MCC performance of defect models by less than 5 percentage points (with a negligible to small effect size), and negligibly impacts the stability of the performance of defect models.

In summary, we find that (1) correlated metrics impact the ranking of the highest ranked metric that is produced by the 9 studied model interpretation techniques of logistic regression and random forest. On the other hand, we find that removing correlated metrics (2) improves the consistency of the highest ranked metric regardless of how a model is specified (except for ANOVA Type-I which is sensitive to the exact specification of a model); (3) improves the consistency of the highest ranked metric among the studied interpretation techniques; and (4) does not substantially decrease the AUC, F-measure, MCC performance, and stability of defect models, suggesting that the benefits of removing correlated metrics outweigh the costs.

*Based on our results, future studies must (1) mitigate (e.g., remove) correlated metrics prior to constructing a defect model (in particular when planing to interpret the model); and (2) avoid using ANOVA Type-I even if all correlated metrics are removed. Due to the variety of the built-in interpretation techniques and their settings, <u>our paper highlights the essential need for future studies to report the exact specification of their models and settings of the used interpretation techniques.</u>*

## 1.1 Novelty Statements

This paper is the first to present:
(1) An exploratory study of the nature of correlated metrics in commonly-studied defect datasets throughout our community (Section 3).
(2) An investigation of the impact of correlated metrics on the consistency of the produced rankings by the interpretation techniques (RQ1).
(3) An empirical evaluation of the consistency of such rankings after removing all correlated metrics (RQ2, RQ3).
(4) An investigation of the impact of removing all correlated metrics on the performance and stability of defect models (RQ4).

## 1.2 Paper Organization

Section 2 describes the studied correlation analysis approaches, commonly used analytical learners, and interpretation techniques. Section 3 presents the results of an exploratory study of the nature of correlated metrics in defect datasets and its impact on importance scores of metrics. Section 4 discusses the design of our case study, while Section 5 presents our results with respect to our four research questions. Section 6 provides practical guidelines for future studies. Section 7 discusses the threats to the validity of our study. Finally, Section 8 draws conclusions.

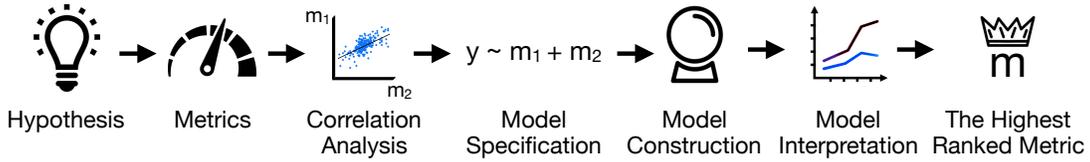

Figure 1: An overview of the analytical modelling process.

Table 1: A summary of the studied correlation analysis approaches, the two studied analytical learners, and the 9 studied interpretation techniques.

| Correlation Analysis | Analytical Learner | Interpretation Technique | Test Statistic | R function |
|---|---|---|---|---|
| Variable Clustering [62, 87–89] | Logistic Regression (`glm` and `lrm`) [7, 8, 64, 65, 95] | Type-I | Deviance | `stats::anova(glm.model)` |
| | | Type-II | Wald | `car::Anova(glm.model, type=2, test.statistic='Wald')` |
| | | | Likelihood Ratio (LR) | `car::Anova(glm.model, type=2, test.statistic='LR')` |
| Variance Inflation Factor [5, 20, 56, 78, 79] | | | F | `car::Anova(glm.model, type=2, test.statistic='F')` |
| | | | Chi-square | `rms::anova(lrm.model, test='Chisq')` |
| Redundancy Analysis [3, 41, 62, 80, 84] | Random Forest [30, 31, 42, 61, 70] | Scaled Gini | MeanDecreaseGini | `randomForest::importance(model, type = 2, scale = TRUE)` |
| | | Non-scaled Gini | MeanDecreaseGini | `randomForest::importance(model, type = 2, scale = FALSE)` |
| | | Scaled Permutation | MeanDecreaseAccuracy | `randomForest::importance(model, type = 1, scale = TRUE)` |
| | | Non-scaled Permutation | MeanDecreaseAccuracy | `randomForest::importance(model, type = 1, scale = FALSE)` |

## 2 BACKGROUND

Figure 1 provides an overview of the commonly-used analytical modelling process. First, one must formulate a set of hypotheses pertaining to phenomena of interest (e.g., whether the size of a module increases the risk associated with that module). Second, one must determine a set of metrics which operationalize the hypothesis of interest (e.g., the total lines of code for size, and the number of field reported bugs to capture the risk that is associated with a module). Third, one must perform a correlation analysis to remove correlated metrics. Forth, one must define a model specification (e.g., the ordering of metrics) to be used when constructing an analytical model. Fifth, one is then ready to construct an analytical model using a machine learning technique (e.g., a random forest model) or a statistical learning technique (e.g., a regression model). Finally, one analyzes the ranking of the metrics using model interpretation techniques (e.g., ANOVA or Breiman's Variable Importance) in order to test the hypotheses of interest.

Based on a literature survey of Hall *et al.* [32] and Shihab [77], we select the commonly-used correlation analysis approaches: variable clustering analysis (VarClus), variance inflation factor (VIF), and redundancy analysis (Redun). The aforementioned serveys guide our selection of the two commonly-used analytical learners: logistic regression [7, 8, 20, 47, 60, 64, 65, 71, 95] and random forest [30, 31, 42, 61, 70]. These techniques are two of the most commonly-used analytical learners for defect models and they have built-in techniques for model interpretation (i.e., ANOVA for logistic regression and Breiman's Variable Importance for random forest). Finally, we select 9 model interpretation techniques, ANOVA Type-I, ANOVA Type-II with 4 test statistics (i.e., Wald, Likelihood Ratio, F, and Chi-square), scaled and non-scaled Gini Importance, and scaled and non-scaled Permutation Importance. Due to the same way in which the importance of metrics for ANOVA Type-II and Type-III are calculated for an additive model (e.g., $(y \sim m_1 + ... + m_n)$), we only evaluate ANOVA Type-II. Table 1 provides a summary of the studied correlation analysis approaches, the two studied analytical learners, and the 9 studied interpretation techniques.

### 2.1 Correlation Analysis

**Variable Clustering (VarClus)** is a hierarchical clustering view of the correlation between metrics [74]. We use the implementation of the variable clustering analysis as provided by the `varclus` function of the `Hmisc` R package [34], which is made up of 2 steps.

*(Step 1) Compute the correlations between metrics.* We use the Spearman rank correlation test ($\rho$) to assess the correlation between metrics. We choose the Spearman test instead of other types of correlation (e.g., Pearson) because the Spearman test is resilient to non-normality in a dataset as commonly present in software engineering and defect datasets, in particular.

*(Step 2) Select one metric from each of the sub-hierarchies for inclusion in a model.* Once, a hierarchical overview of the correlation among metrics is constructed, we use the interpretation of correlation coefficients ($|\rho|$) as provided by Kraemer *et al.* [49], i.e., a correlation coefficient of above 0.7 is considered a strong correlation. Thus, for each sub-hierarchy of software metrics with a correlation $|\rho| > 0.7$, we select only one metric from the sub-hierarchy for inclusion in our models. As suggested by prior studies [55, 57, 88], we select the simplest metric to calculate (or interpret) for each sub-hierarchy.

While the variable clustering analysis (VarClus) technique reduces collinearity among metrics, it does not detect all of the inter-correlated metrics (a.k.a. *multi-collinearity*), i.e., a metric that can be predicted from the other metrics in the model with a certain degree of accuracy.

**Variance Inflation Factor (VIF)** measures the magnitude of multi-collinearity [25]. We use the implementation of the Variance Inflation Factor analysis as provided by the `vif` function of the `rms` R package [36]. Broadly speaking, VIF is made up of 3 steps.

*(Step 1) Construct a regression model for each metric.* For each metric, we construct a model using the other metrics to predict that particular metric.

*(Step 2) Compute a VIF score for each metric.* The VIF score for each metric is computed using the following formula: $\text{VIF} = \frac{1}{1-R^2}$, where $R^2$ is the explanatory power of the regression model from Step 1. A higher VIF score indicates that a given metric can be accurately predicted by the other metrics. Thus, that given metric is considered redundant and should be removed from our model.

*(Step 3) Remove metrics with a VIF score that is higher than a given threshold.* We remove metrics with a VIF score that is higher than a given threshold. We use a VIF threshold of 5 to determine the magnitude of multi-collinearity, as it is commonly used in prior work [5, 24, 56, 78, 79]. Similar to the variable clustering analysis, we repeat the above three steps until the VIF scores of all remaining metrics are lower than the threshold.

**Redundancy Analysis (Redun)** shares the same computational approach with the Variance Inflation Factor (VIF). Instead of using a VIF score (i.e., an inverse proportion of the $R^2$), Redun uses the $R^2$ as a threshold to determine inter-correlated metrics. Thus, we exclude the redundancy analysis from our evaluation.

## 2.2 Analytical Learners

**Logistic regression** is a statistical learner which explains the relationship between one binary dependent variable (e.g., defect-proneness) and one or more independent variables (e.g., software metrics).

**Random forest** is a machine learner that constructs multiple decision trees from bootstrap samples [10]. The final predicted class of a software module is the aggregation of the votes from all of the constructed trees.

## 2.3 Interpretation Techniques

### 2.3.1 Analysis of Variance for Logistic Regression

Analysis of Variance (a.k.a. multi-way ANOVA) is a statistical test that examines the importance of multiple independent variables (e.g., two or more software metrics) on the outcome (e.g., defect-proneness) [23]. The significance of each metric in a regression model is estimated from the calculation of the Sum of Squares (SS)—i.e., the explained variance of the observations with respect to their mean value. There are two commonly-used approaches to calculate the Sum of Squares for ANOVA, namely, Type-I and Type-II. We provide a description of the three types of ANOVA below.

**Type-I**, *one of the most commonly-used interpretation techniques and the default interpretation technique for a logistic regression (`glm`) model in R*, examines the importance of each metric in a sequential order [16, 24]. In other words, Type-I measures the improvement of the Residual Sum of Squares (RSS) (i.e., the unexplained variance) when each metric is sequentially added into the model. Hence, Type-I attributes as much variance as it can to the first metric before attributing residual variance to the second metric in the model specification. Thus, the importance (i.e., produced ranking) of metrics is dependent on the ordering of metrics in the model specification.

The calculation starts from the RSS of the preliminary model ($y \sim 1$), i.e., a null model that is fitted without any software metrics. We then compute the RSS of the first metric by fitting a regression model with the first metric ($y \sim m_1$). Thus, the importance of the first metric ($m_1$) is the improvement between the unexplained variances (RSS) of the preliminary model and the model that is constructed by the first metric.

$$\text{SS}(m_1) = \text{RSS}(\text{Model}_{\text{null}}) - \text{RSS}(m_1) \quad (1)$$

Similar to the computation of the importance of the first metric, the importance of the remaining metrics is computed using the following equation.

$$\text{SS}(m_i) = \text{RSS}(m_1 + ... + m_{i-1}) - \text{RSS}(m_1 + ... + m_i) \quad (2)$$

**Type II**, an enhancement to the ANOVA Type-I, examines the importance of each metric in a hierarchical nature, i.e., the ordering of metrics is rearranged for each examination [16, 24]. The importance of metrics (Type-II) measures the improvement of the Residual Sum of Squares (RSS) (i.e., the unexplained variance) when adding a metric under examination to the model after the other metrics. In other words, the importance of metrics (Type-II) is equivalent to a Type-I where a metric under examination appears at the last position of the model. The intuition is that the Type-II is evaluated after all of the other metrics have been accounted for. The importance of each metric (i.e., $\text{SS}(m_e)$) measures the improvement of the RSS of the model that is constructed by adding only the other metrics except the metric under examination, and the RSS of the model that is constructed by adding the other metrics where the metric under examination appears at the last position of the model. For example, given a set of $M$ metrics, and $e, i, j \in [1, M]$, the importance of each metric $m_e$ can be explained as follows:

$$\text{SS}(m_e) = \text{RSS}(m_i + ... + m_j) - \text{RSS}(m_i + ... + m_j + m_e) \quad (3)$$

where $m_e$ is the metric under examination and $m_i + ... + m_j$ is a set of the other metrics except the metric under examination.

In this paper, we consider different variants of test statistics for ANOVA Type-II (i.e., Wald, Likelihood Ratio (LR), F, and Chi-square).

### 2.3.2 Variable Importance for Random Forest

Variable importance (a.k.a. VarImp) is an approach to examine the importance of software metrics for random forest classifiers. There are two commonly-used calculation approaches of variable importance scores, namely, Gini Importance and Permutation Importance, which we describe below.

**Gini Importance (a.k.a. MeanDecreaseGini)** determines the importance of metrics from the decrease of the Gini Index, i.e., the distinguishing power for the defective class due to a given metric [10, 11]. We start from a random forest model that is constructed using the original dataset with multiple trees, where each tree is constructed using a bootstrap sample. For each tree, a parent node (i.e., $G_{\text{Parent}}$) is splitted by the best cut-point into two descendent nodes (i.e., $G_{\text{Desc.1}}$ and $G_{\text{Desc.2}}$). The calculation of the Gini Importance for each metric is made up of 2 steps:

*(Step 1) Compute the DecreaseGini for all of the trees in the random forest model.* The DecreaseGini is the improvement of the ability to distinguish between two classes across parent and its descendent nodes. We compute the DecreaseGini using the following equation:

$$\text{DecreaseGini}(m_i) = I_{m_i} = G_{\text{Parent}} - G_{\text{Desc.1}} - G_{\text{Desc.2}} \quad (4)$$

where $G$ is the Gini Index, i.e., the distinguishing power of defective class for a given metric. The Gini Index is computed using the following equation: $G = \sum_{i=1}^{N_{\text{Class}}} p_i(1 - p_i)$, where $N_{\text{Class}}$ is the number of classes and $p_i$ is the proportion of Class$_i$.

*(Step 2) Compute the MeanDecreaseGini measure.* Finally, the importance for each metric (i.e., MeanDecreaseGini) is the average of the DecreaseGini values from all of the splits of that metric across all the trees in the random forest model.

In this paper, we consider both the scaled and non-scaled importance scores for the Gini Importance.

**Permutation Importance (a.k.a. MeanDecreaseAccuracy)** determines the importance of metrics from the decrease of the accuracy (i.e., the misclassification rate) when the values of a given metric are randomly permuted [10, 11]. Similar to MeanDecreaseGini, we start from a random forest model that is constructed using an original dataset with multiple trees, where each tree is constructed using an out-of-sample bootstrap. The calculation of the Permutation Importance is made up of 2 steps:

*(Step 1) Compute the DecreaseAccuracy of each tree in the random forest model.* The DecreaseAccuracy is the decrease of the accuracy (i.e., misclassification rate) between a model that is tested using the original out-of-bag testing samples and a model that is tested using permuted out-of-bag testing samples, i.e., a dataset with one metric permuted, while all other metrics are unchanged).

*(Step 2) Compute the MeanDecreaseAccuracy measure.* Finally, the importance for each metric (i.e., MeanDecreaseAccuracy) is the average of the DecreaseAccuracy values across all of the trees in the random forest model.

Similar to Gini Importance, we consider both the scaled and non-scaled importance scores for the Permutation Importance.

## 3 MOTIVATING ANALYSES

In this section, we perform motivating analyses to investigate (1) the prevalence of correlated metrics in defect datasets, (2) the impact of the number of correlated metrics on the importance scores of metrics, and (3) the impact of the ordering of correlated metrics in a model specification on the importance ranking of metrics.

### 3.1 The prevalence of correlated metrics in defect datasets

**Approach**. To ensure that the studied metrics are of importance to practitioners when interpreting defect models, we only focus on the correlated metrics that share a strong relationship with defect-proneness. We perform the following steps for each dataset from a collection of the 101 publicly-available defect datasets (see Section 4.1).

*(Step 1) Analyze the relationship between each of the metrics and defect-proneness.* We first identify the software metrics that share a strong relationship with defect-proneness. To analyze the relationship for each of the metrics and defect-proneness, similar to prior work [87, 89], we use the Cliff's $|\delta|$ effect size to compute the magnitude of the difference of the metric values between defective and clean modules. We use a set of thresholds that are proposed by Romano *et al.* [73], i.e., "negligible" for $|\delta| < 0.147$, "weak" for $|\delta| < 0.33$, "medium" for $|\delta| < 0.474$, otherwise "strong". The intuition is that a larger difference of values between defective and clean modules would present an effective software metric in defect models. We use the implementation of the Cliff's $|\delta|$ effect size as provided by the `cliff.delta` function of the `effsize` R package [90].

*(Step 2) Analyze the correlation among software metrics.* We apply a variable clustering analysis (VarClus) to examine the correlations among software metrics (see Section 2.1). Then, we identify a dataset where correlated metrics share a strong relationship with defect-proneness.

**Results**. **Correlated metrics that share a strong relationship with defect-proneness are prevalent in 83 of the 101 (82%) publicly-available defect datasets.** In addition, we observe that there are 1-8 clusters of correlated metrics, where the size of a cluster ranges from 2 to 21 metrics. For example, we find that 16 of 21 metrics (76%) in the `kc2` dataset are highly-correlated with each other. We also find that correlated metrics within a cluster share the same magnitude of the relationship between each of the metrics and defect-proneness. Therefore, any one of the correlated metrics within a cluster can be selected as a representative metric of a cluster.

### 3.2 The impact of the number of correlated metrics on the importance scores of metrics in a model

**Approach**. To assess the impact of correlated metrics on the importance scores of metrics in a model, we analyze the importance scores of each metric in a model when correlated metrics are included in the model. We select the `eclipse-2.0` dataset as the subject of our analysis, since it is widely used in a large number of defect prediction studies [6, 63, 94]. We start from a *mitigated dataset*, i.e., a dataset where correlated metrics are removed. To mitigate correlated metrics, we apply both the variable clustering analysis (VarClus) and the variance inflation factor analysis (VIF) (see Section 2.1). To demonstrate the impact of the number of correlated metrics on the importance scores of metrics, we sequentially add correlated metrics that are associated with the one metric that shares the strongest relationship with defect-proneness. We estimate the magnitude of the relationship using the Cliff's $|\delta|$ estimate. We find that `TLOC` is the metric that shares the strongest relationship with defect-proneness for the `eclipse-2.0` dataset. The variable clustering analysis (VarClus) shows that there are 12 metrics that are highly-correlated with the `TLOC` metric. Then, we examine the importance scores of TLOC for each of the models where metrics that are correlated with TLOC are sequentially added at a time to the first position of the model specification. We measure the importance scores of TLOC using the Type-I, Type-II (LR) for logistic regression models, and the Gini Importance, and scaled and non-scaled Permutation Importance techniques for random forest models. We then plot the percentage relative difference of the importance scores of the TLOC of models that are constructed using mitigated datasets and models with correlated metrics.

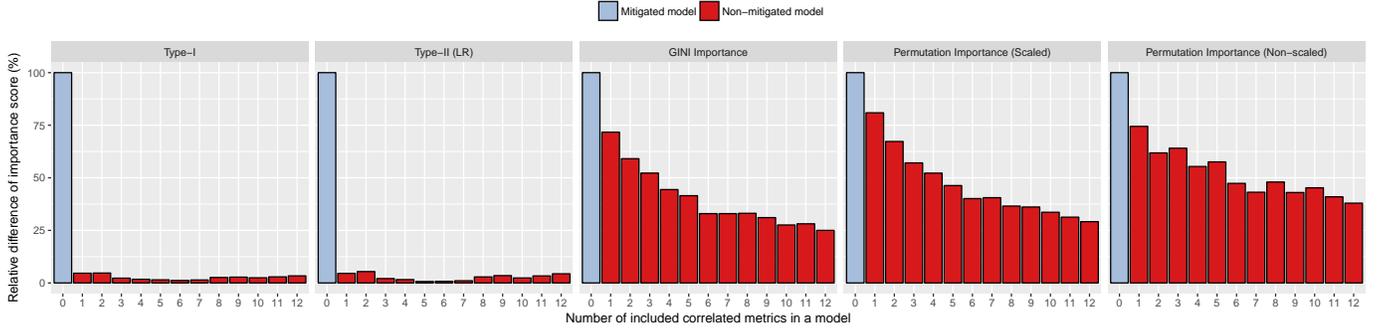

Figure 2: The relative difference of the importance scores of the TLOC when correlated metrics are included in the model in comparison to a model where correlated metrics are not included. The x-axis value of 0 (i.e., a light blue bar) represents the mitigated model (i.e., all correlated metrics are removed), while the x-axis values of 1-12 (i.e., red bars) represent the number of included correlated metrics to TLOC in a non-mitigated model.

**Results**. **Irrespective of the interpretation technique, the importance scores substantially decrease when there are correlated metrics in the models.** Figure 2 shows that the importance scores relatively decrease by 95%, 95%, 28%, 19%, and 26% for the Type-I, Type-II (LR), Gini Importance, and scaled and non-scaled Permutation Importance techniques, respectively. This finding suggests that the importance scores of Type-I, Type-II (LR), Gini Importance, and Permutation Importance techniques tend to inflate (or deflate) when correlated metrics exist in a model specification.

### 3.3 The impact of the ordering of correlated metrics in a model specification on importance scores of such metrics

**Approach**. To assess the impact of the ordering of correlated metrics in a model specification on importance scores, we analyze the importance scores of a defect model when the ordering of correlated metrics in a model specification are rearranged. Similar to prior analysis (Section 3.2), we use the eclipse-2.0 dataset as the subject of our analysis. Since the variable clustering analysis of Section 3.2 shows that there are 12 metrics that are highly-correlated with TLOC, we randomly select 5 of the 12 metrics (i.e., TLOC, MLOC_sum, FOUT_sum, VG_sum, and NBD_sum) in order to simplify the demonstration. We start from a dataset with the 5 selected correlated metrics. We then construct defect models using logistic regression models and random forest models. We examine the importance scores of the 5 metrics using Type-I and Gini Importance (see Section 2.3) with two model formulas, i.e., TLOC appears at the first position for a model, while TLOC appears at the last position for the other model.

Table 2: The percentage of the importance scores of Type-I for logistic regression models (LR) and Gini Importance for random forest models (RF).

|  |  | Specification 1 |  |  | Specification 2 |  |
| --- | --- | --- | --- | --- | --- | --- |
|  |  | Type-I of LR | Gini of RF |  | Type-I of LR | Gini of RF |
| **Metrics** |  | AUC=0.80 | AUC=0.78 |  | AUC=0.80 | AUC=0.78 |
| TLOC | [1] | 97.55%*** | 24.33% | [5] | 1.75%*** | 22.38% |
| MLOC_sum | [2] | 0.07% | 21.68% | [1] | 93.12%*** | **22.57%** |
| FOUT_sum | [3] | 0.12% | 19.61% | [2] | 0% | 19.29% |
| VG_sum | [4] | 2.23%*** | 20.05% | [3] | 4.58%*** | 21.08% |
| NBD_sum | [5] | 0.03% | 14.32% | [4] | 0.55% | 14.69% |

Statistical importance of deviance according to Chi-square test:
∘ $p \geq .05$; * $p < .05$; ** $p < .01$; *** $p < .001$
The bracketed values indicate the position of a metric in the model specification.
The bold text indicates a metric with the highest score.

**Results**. **The importance scores of Type-I and Gini Importance are sensitive to the ordering of correlated metrics in a model specification.** Table 2 shows that TLOC is the highest ranked metric when TLOC is at the first position of the logistic regression and random forest models that are measured using Type-I and Gini Importance. Conversely, TLOC is among the lowest ranked metric when TLOC is at the last position of the logistic regression model that is measured using Type-I. This finding indicates that the importance scores produced by Type-I incorrectly implies that only the correlated metric at the first position in a model specification is important, even though all of the 5 correlated metrics are of similar importance (i.e., they share the same magnitude of the relationship with defect-proneness). Similarly, TLOC is the second-highest ranked metric when TLOC is at the last position of the random forest model that is measured by Gini Importance. We also find that the importance scores are distributed evenly among all of the 5 correlated metrics, since our prior motivating analysis (Figure 2) shows that the importance score is getting diluted as additional correlated metrics are included in the model specification.

> **Summary.** *Our motivating analyses highlight that correlated metrics substantially change the importance scores in both logistic regression and random forest models.*

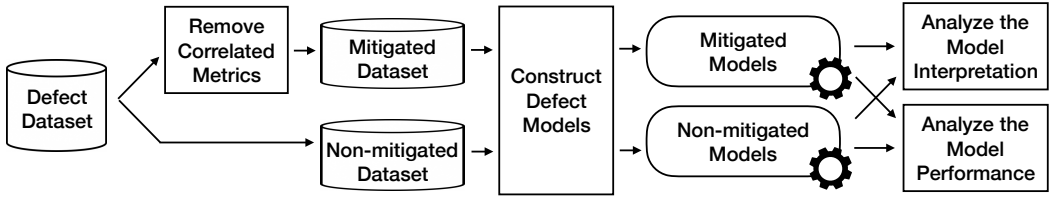

Figure 3: An overview diagram of the design of our case study.

## 4 CASE STUDY DESIGN

In this section, we discuss (1) our criteria for selecting the studied datasets; and (2) the design of the case study that we perform in order to address our four research questions. Figure 3 provides an overview of the design of our case study.

### 4.1 Studied Datasets

In selecting the studied datasets, we identify three important criteria that need to be satisfied:

**Criterion 1—Publicly-available defect datasets.** Prior work raises concerns about the replicability of software engineering studies [72]. In order to foster future replication of our work, we focus on publicly-available defect datasets.

**Criterion 2—Datasets with correlated metrics that have a strong relationship with defect-proneness.** Correlated metrics that have a weak relationship with defect-proneness may not be as important as metrics that have a strong relationship with defect-proneness. To ensure that the studied metrics are of importance to practitioners when interpreting defect models, we only focus on the correlated metrics that share a strong relationship with defect-proneness.

**Criterion 3—Datasets where we can accurately derive interpretations.** Analysts would only consider models that fit the data well (i.e., AUC > 0.7) and stable models (i.e., EPV > 10) [85]. Hence, we only focus on datasets that produce such accurate and stable models.

To satisfy criterion 1, similar to prior work [83], we begin our study using a collection of the 101 publicly-available defect datasets that are collected from 5 different corpora, i.e., 76 datasets from the Tera-PROMISE Repository, 12 clean NASA datasets as provided by Shepperd *et al.* [76], 5 datasets as provided by Kim *et al.* [45, 92], 5 datasets as provided by D'Ambros *et al.* [18, 19], and 3 datasets as provided by Zimmermann *et al.* [95]. To satisfy criterion 2, we exclude 28 datasets where their correlated metrics do not share a strong relationship with defect-proneness. To satisfy criterion 3, we exclude 64 datasets with an EPV value below 10 and 4 datasets on which models that are constructed produce an AUC value below 0.7. Hence, we focus on 15 datasets of systems that span across proprietary and open-source systems. Table 3 shows a statistical summary of the studied datasets.

### 4.2 Remove Correlated Metrics

To investigate the impact of correlated metrics on the performance and interpretation of defect models and address our four research questions, we start by removing highly-correlated metrics in order to produce mitigated datasets, i.e., datasets where correlated metrics are removed. To do so, we apply variable clustering analysis (VarClus) and variable influence factor analysis (VIF) (see Section 2.1). We use the Spearman correlation ($|\rho|$) threshold of 0.7 to identify correlated metrics. We use a VIF threshold of 5 to identify inter-correlated metrics. We use the implementation of the variable clustering analysis as provided by the `varclus` function of the `Hmisc` R package [34]. We use the implementation of the VIF analysis as provided by the `vif` function of the `rms` R package [36].

Table 3: A statistical summary of the studied datasets.

| Project | Dataset | Modules | Metrics | Correlated Metrics | EPV | $AUC_{LR}$ | $AUC_{RF}$ |
|---|---|---|---|---|---|---|---|
| Apache | Lucene 2.4 | 340 | 20 | 9 | 10 | 0.74 | 0.77 |
| | POI 2.5 | 385 | 20 | 11 | 12 | 0.80 | 0.90 |
| | POI 3.0 | 442 | 20 | 10 | 14 | 0.79 | 0.88 |
| | Xalan 2.6 | 885 | 20 | 8 | 21 | 0.79 | 0.85 |
| | Xerces 1.4 | 588 | 20 | 11 | 22 | 0.91 | 0.95 |
| Eclipse | Debug 3.4 | 1,065 | 17 | 9 | 15 | 0.72 | 0.81 |
| | JDT | 997 | 15 | 10 | 14 | 0.81 | 0.82 |
| | Mylyn | 1,862 | 15 | 10 | 16 | 0.78 | 0.74 |
| | PDE | 1,497 | 15 | 9 | 14 | 0.72 | 0.72 |
| | Platform 2.0 | 6,729 | 32 | 24 | 30 | 0.82 | 0.84 |
| | Platform 3.0 | 10,593 | 32 | 24 | 49 | 0.79 | 0.81 |
| | SWT 3.4 | 1,485 | 17 | 7 | 38 | 0.87 | 0.97 |
| NASA | PC5 | 1,711 | 38 | 27 | 12 | 0.73 | 0.78 |
| Proprietary | Prop 1 | 18,471 | 20 | 10 | 137 | 0.75 | 0.79 |
| | Prop 4 | 8,718 | 20 | 11 | 42 | 0.74 | 0.72 |

### 4.3 Construct Defect Models

To examine the impact of correlated metrics on the performance and interpretation of defect models, we construct our models using the *non-mitigated datasets* (i.e., datasets where correlated metrics are not removed) and *mitigated datasets* (i.e., datasets where correlated metrics are removed). To construct defect models, we perform the following steps:

**(CM1) Generate bootstrap samples.** To ensure that our conclusions are statistically sound and robust, we use the out-of-sample bootstrap validation technique, which leverages aspects of statistical inference [22, 26, 35, 85]. We first generate bootstrap sample of sizes $N$ with replacement from the mitigated and non-mitigated datasets. The generated sample is also of size $N$. We construct models using the bootstrap samples, while we measure the performance of the models using the samples that do not appear in the bootstrap samples. On average, 36.8% of the original dataset will not appear in the bootstrap samples, since the samples are drawn with replacement [22]. We repeat the out-of-sample bootstrap process for 100 times and report their average performance.

**(CM2) Construct defect models.** For each bootstrap sample, we construct logistic regression and random forest models. We use the implementation of logistic regression as provided by the `glm` function of the `stats` R package[86] and the `lrm` function of the `rms` R package[36]. We use the implementation of random forest as provided by the `randomForest` function of the `randomForest` R package [12].

### 4.4 Analyze the Model Interpretation

To address RQ1, RQ2, and RQ3, we analyze the importance ranking of metrics of the models that are constructed using non-mitigated datasets and mitigated datasets. The analysis of model interpretation is made up of 2 steps.

**(MI1) Compute the importance score of metrics.** We investigate the impact of correlated metrics on the interpretation of defect models using different model interpretation techniques. Thus, we apply the 9 studied model interpretation techniques, i.e., Type-I, Type-II (Wald, LR, F, Chisq), scaled and non-scaled Gini Importance, and scaled and non-scaled Permutation Importance. The technical description and implementation details of the studied model interpretation techniques are provided in Section 2.3 and Table 1.

**(MI2) Identify the highest ranked metric.** To statistically identify the highest ranked metric, we apply the improved Scott-Knott Effect Size Difference (ESD) test (v2.0) [82]. The Scott-Knott ESD test is a mean comparison approach that leverages a hierarchical clustering to partition a set of treatment means (i.e., means of importance scores) into statistically distinct groups *with statistically non-negligible difference*. The Scott-Knott ESD test ranks each metric at only a single rank, however several metrics may appear within one rank. Finally, we identify the highest ranked metric for the non-mitigated and mitigated models. Thus, each metric has a rank for each model interpretation technique and for each of the mitigated and non-mitigated models. We use the implementation of Scott-Knott ESD test as provided by the `sk_esd` function of `ScottKnottESD` R package [82].

### 4.5 Analyze the Model Performance

To address RQ4, we analyze the performance of the models that are constructed using non-mitigated datasets and mitigated datasets.

First, we use the Area Under the receiver operator characteristic Curve (AUC) to measure the discriminatory power of our models, as suggested by recent research [28, 52, 69]. The AUC is a threshold-independent performance measure that evaluates the ability of classifiers in discriminating between defective and clean modules. The values of AUC range between 0 (worst performance), 0.5 (no better than random guessing), and 1 (best performance) [33].

Second, we use the F-measure, i.e, a threshold-independent measure. F-measure is a harmonic mean (i.e., $\frac{2 \cdot \text{precision} \cdot \text{recall}}{\text{precision}+\text{recall}}$) of precision ($\frac{\text{TP}}{\text{TP+FP}}$) and recall ($\frac{\text{TP}}{\text{TP+FN}}$). Similar to prior studies [2, 94], we use the default probability value of 0.5 as a threshold value for the confusion matrix, i.e., if a module has a predicted probability above 0.5, it is considered defective; otherwise, the module is considered clean.

Third, we use the Matthews Correlation Coefficient (MCC) measure, i.e, a threshold-independent measure, as suggested by prior studies [53, 75]. MCC is a balanced measure based on true and false positives and negatives that is computed using the following equation: $\frac{\text{TP} \times \text{TN} - \text{FP} \times \text{FN}}{\sqrt{(\text{TP+FP})(\text{TP+FN})(\text{TN+FP})(\text{TN+FN})}}$.

## 5 CASE STUDY RESULTS

In this section, we present the results of our case study with respect to our four research questions.

### (RQ1) How do correlated metrics impact the interpretation of defect models?

**Motivation**. Prior work raises concerns that metrics are often correlated [29, 39, 50, 51, 84, 93]. For example, Landman *et al.* [50, 51], Herraiz *et al.* [39], and Gil *et al.* [29] point out that code complexity is often correlated with lines of code. Unfortunately, a literature survey of Shihab [77] shows that as much as 63% of prior defect studies do not mitigate (e.g., remove) correlated metrics prior to constructing defect models. Yet, little is known about the impact of correlated metrics on the interpretation of defect models.

**Approach**. To address RQ1, we analyze the difference of the produced importance rankings when including a metric that is correlated with the highest ranked metric in a model. To do so, we start from mitigated datasets (see Section 4.2). We first identify the highest ranked metric for each of the 9 studied interpretation techniques. We use VarClus to select only one of the metrics that is correlated with the highest ranked metric in order to generate non-mitigated datasets. We then append the correlated metric to the first position of the specification of the mitigated models. Thus, the specification for the mitigated models is $y \sim m_{\text{high}} + ...$, while the specification for the non-mitigated models is $y \sim m_c + m_{\text{high}} + ...$, where $m_c$ is the metric that is correlated with the highest ranked metric ($m_{\text{high}}$). For each of the mitigated and non-mitigated datasets, we construct defect models (see Section 4.3) and apply the 9 studied model interpretation techniques (see Section 4.4). For each interpretation technique, we analyze the difference in the ranks of the highest ranked metric of the models that are constructed using the mitigated and non-mitigated datasets. For example, if a metric $m_{\text{high}}$ appears in the top rank in both of the mitigated and non-mitigated models, then the metric would have a rank difference of 0. However, if $m_{\text{high}}$ appears in the third rank in the non-mitigated model, then the rank difference of $m_{\text{high}}$ would be 2. Finally, we compute the percentage of the studied datasets for each difference in the ranks between the highest ranked metric of the models that are constructed using the mitigated and non-mitigated datasets.

**Results**. Figure 4 shows the rank difference of the highest ranked metric of the models that are constructed using the mitigated and non-mitigated datasets for each of the studied interpretation techniques.

**ANOVA Type-I is the most sensitive technique to correlated metrics.** We expect that the highest ranked metric in the mitigated model will remain as the highest ranked metric in the non-mitigated model. Unfortunately, Figure 4 shows that this expectation does not hold true in any of the studied datasets for ANOVA Type-I. We suspect that the impact of correlated metrics on the interpretation of Type-I has to do with the sequential nature of the calculation of the Sum of Squares, i.e., Type-I attributes as much variance as it can to the first metric before attributing residual variance to the second metric in the model specification.

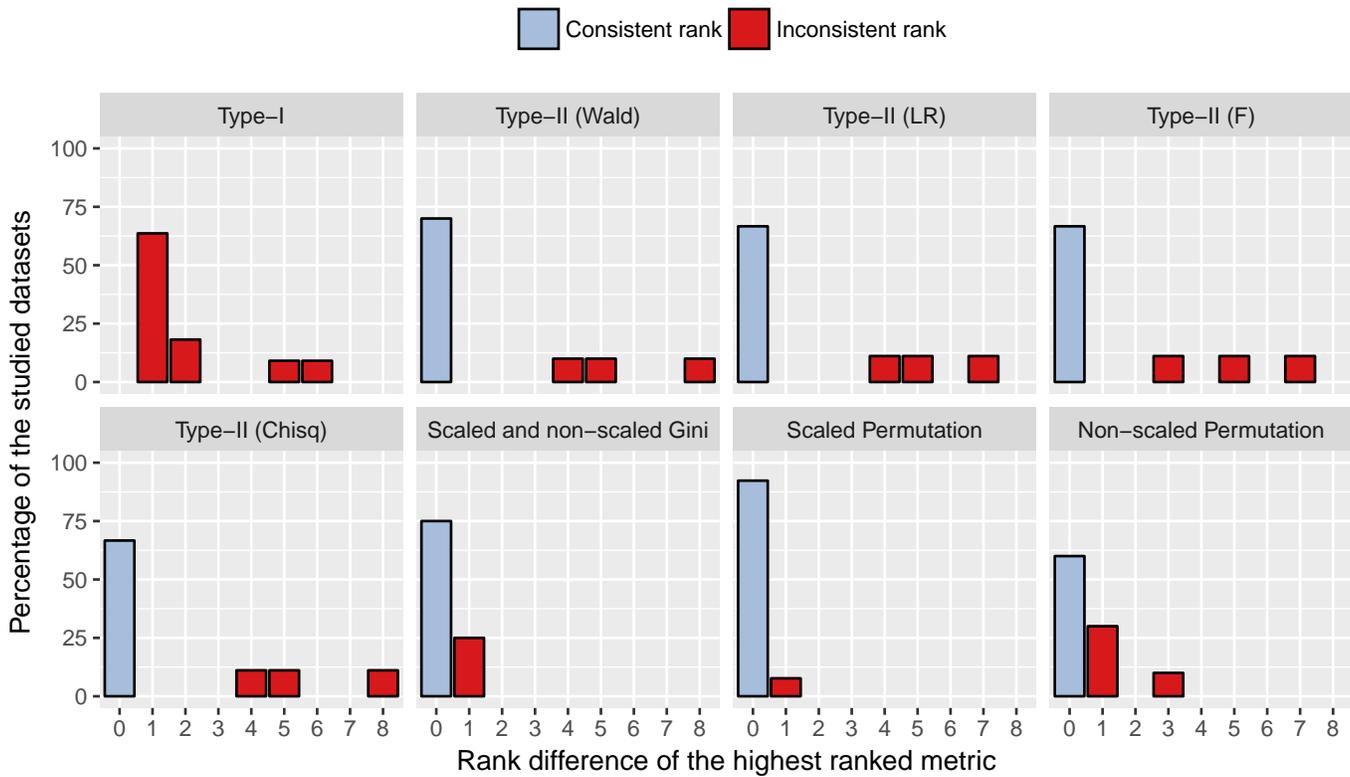

Figure 4: The percentage of the studied datasets for each difference in the ranks between the highest ranked metric of the models that are constructed using the mitigated and non-mitigated datasets. The light blue bars represent the consistent rank of the highest ranked metric between mitigated and non-mitigated models, while the red bars represent the inconsistent rank of the highest ranked metric between mitigated and non-mitigated models.

**On the other hand, correlated metrics tend to have a lower impact on the interpretation of Type-II, Gini Importance, and Permutation Importance.** Fortunately, we find that the highest ranked metric in the mitigated model will remain as the highest ranked metric in the non-mitigated model for 70%, 67%, 67%, 67%, 75%, 92%, and 60% of the studied datasets for ANOVA Type-II (Wald), Type-II (LR), Type-II (F), Type-II (Chisq), Gini Importance, and scaled and non-scaled Permutation Importance, respectively. The impact of correlated metrics on the interpretation of Type-II has to do with the hierarchical nature of the calculation of Sum Squares, i.e., the importance score of the metric is evaluated after all of the other metrics have been accounted for. Moreover, the impact of correlated metrics on the interpretation of the Gini Importance and Permutation Importance has to do with the random process for constructing multiple trees and the calculation of importance scores for a random forest model. First, the random process of random forest may generate some trees that are constructed using correlated metrics. Second, the average importance scores from multiple trees are diluted due to the trees that are constructed using correlated metrics. Although Figure 4 shows that

Gini Importance and Permutation Importance have a lower impact than Type-I when including only one correlated metric in the model, our motivating analysis (see Section 3.2) suggests that the ranking produced by Gini Importance and Permutation Importance will be substantially impacted due to the drastic decrease of the importance scores when including a higher number of correlated metrics.

> *Irrespective of the built-in interpretation techniques for logistic regression and random forest, correlated metrics introduce inconsistency to the ranking of the highest ranked metric, highlighting the risks of not mitigating correlated metrics before constructing models.*

**(RQ2) After removing all correlated metrics, how consistent is the interpretation of defect models among different model specifications?**

**Motivation**. Our motivating analysis (Section 3.3) and the results of RQ1 confirm that the ranking of the highest ranked metric substantially changes when the ordering of correlated metrics in a model specification is rearranged, suggesting that correlated metrics must be removed. However, after removing correlated metrics, little is known if the interpretation of defect models would become consistent when rearranging the ordering of metrics.

**Approach**. To address RQ2, we analyze the ranking of the highest ranked metric of the models that are constructed

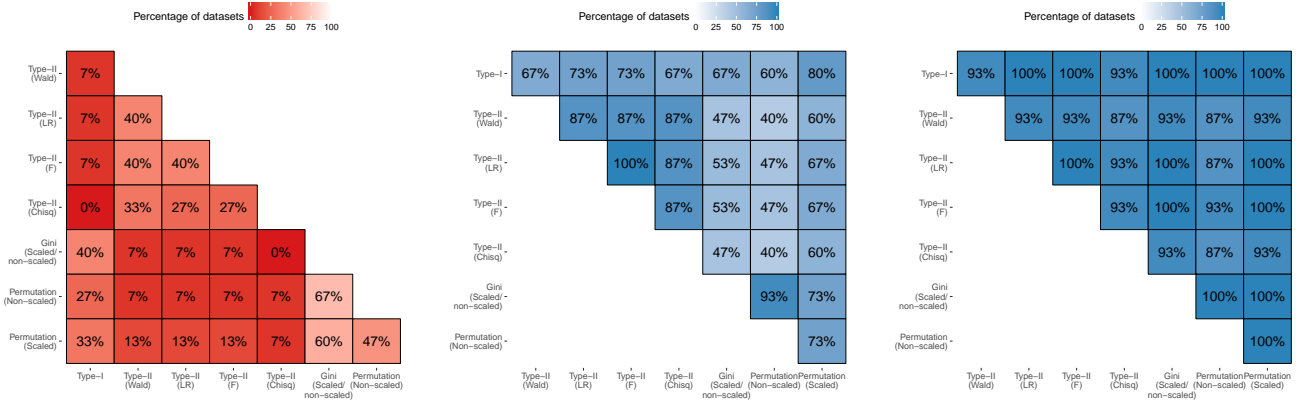

(a) The top-1 ranked metric for non-mitigated models.
(b) The top-1 ranked metric for mitigated models.
(c) The top-3 ranked metrics for mitigated models.

Figure 5: The percentage of datasets where the top-ranked metric is consistent between the two studied model interpretation techniques. While the lower-left side of the matrix (i.e., red shades) shows the percentage before removing correlated metrics, the upper-right side of the matrix (i.e., blue shades) shows the percentage after removing correlated metrics.

using different ordering of metrics from mitigated datasets. To do so, we start from mitigated datasets that are produced by Section 4.2. For each of the datasets, we construct defect models (see Section 4.3) and apply the 9 studied model interpretation techniques (see Section 4.4) in order to identify the highest ranked metric according to each technique. Then, we regenerate the models where the ordering of metrics is rearranged—the highest ranked metric is at each position from the first to the last for each dataset. Finally, we compute the percentage of datasets where the ranks of the highest ranked metric are inconsistent among the rearranged datasets.

**Results**. **After removing correlated metrics, the highest ranked metric according to Type-II, Gini Importance, and Permutation Importance are consistent. However, the highest ranked metric according to Type-I is still inconsistent regardless of the ordering of metrics.** We find that Type-II, Gini Importance, and Permutation Importance produce a stable ranking of the highest ranked metric for all of the studied datasets regardless of the ordering of metrics.

On the other hand, ANOVA Type-I is the only technique that produces an inconsistent ranking of the highest ranked metric. We find that, for 73% of the studied datasets, ANOVA Type-I produces an inconsistent ranking of the highest ranked metric when the ordering of metrics is rearranged. We expect that the consistency of the ranking of the highest ranked metric can be improved by increasing the strictness of the correlation threshold of the variable clustering analysis (VarClus). Thus, we repeat the analysis using stricter thresholds of the variable clustering analysis (VarClus). We use $|\rho|$ thresholds of 0.5, and 0.6. Unfortunately, even if we increase the strictness of the correlation threshold, Type-I produces the inconsistent ranking of the highest ranked metric for 53% and 47% of the studied datasets, for the threshold of 0.5 and 0.6, respectively.

The inconsistent ranking of the highest ranked metric according to Type-I has to do with the sequential nature of the calculation of the Sum of Squares (see Section 2.3). In other words, Type-I attributes the importance scores as much as it can to the first metric before attributing the scores to the second metric in the model specification. Thus, Type-I is sensitive to the ordering of metrics.

> *After removing all correlated metrics, the highest ranked metric according to Type-II, Gini Importance, and Permutation Importance is consistent. However, the highest ranked metric according to Type-I is inconsistent (as the ranking of a metric is impacted by its order in the model specification when analyzed using Type-I, the default analysis for the* `glm` *model in R, which is commonly used in prior studies).*

**(RQ3) After removing all correlated metrics, how consistent is the interpretation of defect models among the studied interpretation techniques?**

**Motivation**. The findings of prior work often rely heavily on one model interpretation technique [30, 41, 61, 62, 80, 84]. Therefore, the findings of prior work may pose a threat to construct validity, i.e., the findings may not hold true if one uses another interpretation technique. Thus, we set out to investigate if the highest ranked metric is consistent among interpretation techniques after removing correlated metrics.

**Approach**. To address RQ3, we start from mitigated datasets that are produced by Section 4.2 and non-mitigated datasets (i.e., the original datasets). We compare the two rankings that are produced from mitigated and non-mitigated models using the 9 interpretation techniques for each of the studied datasets. Then, we compute the percentage of datasets where the highest ranked metric is consistent among the studied model interpretation techniques. Finally, we present the results using a heatmap (as shown in Figure 5) where each cell indicates the percentage of datasets where the highest ranked metric is consistent between the two studied model interpretation techniques. Figures 5a and 5b present the percentage of datasets where the highest ranked metric (i.e., top-1) is consistent between the two studied model interpretation techniques for non-mitigated models and mitigated models, respectively. On the other hand, Figure 5c presents the percentage of datasets where at least one metric

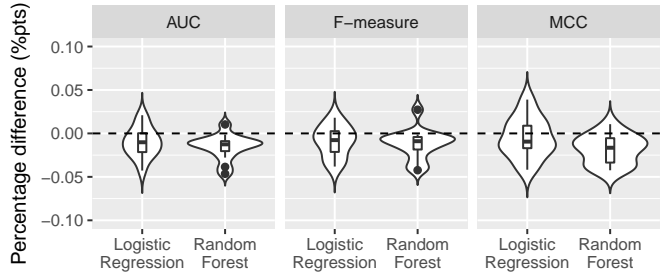
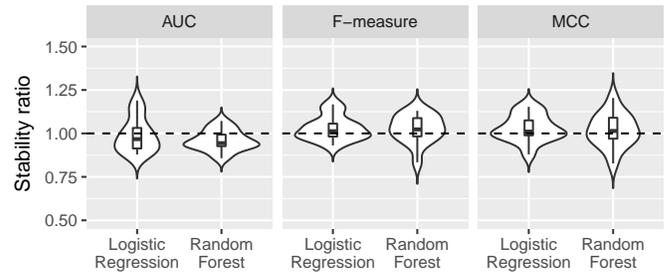

Figure 6: The distributions of the performance difference (% pts) of models that are constructed using non-mitigated and mitigated datasets.

Figure 7: The distributions of the stability ratio of models that are constructed using non-mitigated and mitigated datasets.

in the top-3 ranked metrics is consistent between the two studied model interpretation techniques.

**Results**. **Before removing all correlated metrics, we find that the studied model interpretation techniques do not tend to produce the same highest ranked metric.** According to the lower-left side of the matrix of the Figure 5a, we find that, before removing correlated metrics, the highest ranked metric of Type-II (Chisq) of logistic regression and Gini Importance of random forest is inconsistent for all of the studied datasets. On the other hand, among the variants of the Type-II techniques, the highest ranked metric is consistent for 27%-40% of the studied datasets.

**After removing all correlated metrics, we find that at least one metric in the top-3 ranked metrics of the studied model interpretation techniques is consistent for 87%-100% of the studied datasets.** According to the upper-right side of the matrix of the Figure 5b, we find that, after removing correlated metrics, the highest ranked metric of Type-II (Chisq) and Gini Importance is improved from 0% to 47% of the studied datasets. Moreover, among the variants of the Type-II techniques, the consistency of the highest ranked metric is improved from 27%-40% to 87%-100% of the studied datasets. Most importantly, Figure 5c shows that at least one metric in the top-3 ranked metrics for all of the studied model interpretation techniques is consistent for 87%-100% of the studied datasets. This finding highlights the benefits of removing correlated metrics on the interpretation of defect models—the conclusions of studies that rely on one interpretation technique may not pose a threat after mitigating (e.g., removing) correlated metrics.

> *After removing all correlated metrics, we find that at least one metric in the top-3 ranked metrics of the studied model interpretation techniques is consistent for 87%-100% of the studied datasets, highlighting the benefits of removing all correlated metrics on the interpretation of defect models, i.e., the conclusions of studies that rely on one interpretation technique may not pose a threat after mitigating (e.g., removing) correlated metrics.*

### (RQ4) Does removing all correlated metrics impact the performance and stability of defect models?

**Motivation**. The results of RQ1 show that correlated metrics have a negative impact on the interpretation of defect prediction models, while the results of RQ2 and RQ3 show the benefits of removing correlated metrics on the interpretation of defect models. Thus, removing correlated metrics is highly recommended. However, removing correlated metrics may pose a risk to the performance and stability of defect models. Yet, little is known if removing such correlated metrics decreases the performance and stability of defect models.

**Approach**. To address RQ4, we analyze (1) the AUC, F-measure, and MCC performance difference, and (2) the stability ratio (i.e., a standard deviation of the performance estimates that are produced by the 100 iterations of the out-of sample bootstrap) of models that are constructed using non-mitigated and mitigated datasets. We then measure the Cliff's $|\delta|$ effect size of the magnitude of the difference between (1) the performance, and (2) the stability ratio of the non-mitigated and mitigated models.

**Results**. **Removing all correlated metrics decreases the AUC, F-measure, and MCC performance of defect models by less than 5 percentage points.** Figure 6 shows that the distributions of the performance difference of the models that are constructed using non-mitigated and mitigated datasets are centered at zero. In addition, our Cliff's $|\delta|$ effect size test confirms that the differences between the models that are constructed using mitigated and non-mitigated datasets are negligible to small for the AUC, F-measure, and MCC measures.

**Removing all correlated metrics negligibly impacts the stability of the performance of defect models.** Figure 7 shows that the distributions of the stability ratio of the models that are constructed using non-mitigated and mitigated datasets are centered at one (i.e., there is little difference in model stability after removing all correlated metrics). Moreover, our Cliff's $|\delta|$ effect size test confirms that the difference of the stability ratio between the models that are constructed using mitigated and non-mitigated datasets is negligible.

The negligible to small difference of the performance and stability of models after removing correlated metrics has to do with (1) the strong correlation among the correlated metrics, and (2) the same magnitude of the relationship between each of the correlated metrics and defect-proneness. Thus, the finding confirms our suggestion in Section 3.1 that selecting one of the correlated metrics as a representative metric for each cluster of correlated metrics while removing the other correlated metrics negligibly impacts the perfor-mance and stability of defect models. This finding suggests that the benefits of removing all correlated metrics outweigh the costs—the performance and stability of defect models do not drop drastically when removing correlated metrics.

> *Removing all correlated metrics decreases the AUC, F-measure, and MCC performance of defect models by less than 5 percentage points (with a negligible to small effect size), and negligibly impacts the stability of the performance of defect models.*

## 6 PRACTICAL GUIDELINES

In this section, we offer practical guidelines for future studies:

**(1) Researchers must mitigate (e.g., remove) correlated metrics prior to constructing a defect model**, since RQ1 shows that (1) correlated metrics impact the ranking of metrics according to the studied model interpretation techniques. On the other hand, the results of RQ2, RQ3, and RQ4 show that removing all correlated metrics (2) improves the consistency of the highest ranked metric regardless of the ordering of metrics; (3) improves the consistency of the highest ranked metric among to the studied interpretation techniques; and (4) does not substantially decrease the AUC, F-measure, MCC performance, and stability of defect models, suggesting that the benefits of removing all correlated metrics outweigh the cost.

**(2) Researchers must avoid using ANOVA Type-I even if all correlated metrics are removed**, since RQ2 shows that Type-I produces an inconsistent ranking of the highest ranked metric when the orders of metrics are rearranged, indicating that Type-I is sensitive to the ordering of metrics even when removing all correlated metrics.

## 7 THREATS TO VALIDITY

We now discuss threats to the validity of our study.

### 7.1 Construct Validity

In this work, we only construct regression models in an additive fashion ($y \sim m_1 + ... + m_n$), since metric interactions (i.e., the relationship between each of the two interacting metrics depends on the value of the other metrics) (1) are rarely explored in software engineering; (2) must be statistically insignificant (e.g., absence) for ANOVA Type-II test [16, 24]; and (3) are not compatible with random forest [10] which is one of the most commonly-used analytical learners in software engineering. On the other hand, the importance score of the metric produced by ANOVA Type-III is evaluated after all of the other metrics and all interactions of the metric under examination have been accounted for. Thus, if interactions are significantly present, one should use ANOVA Type-III and avoid using ANOVA Type-II. Due to the same way in which the importance of metrics for ANOVA Type-II and Type-III are calculated in a hierarchical nature (see Section 2.3) for an additive model, we would like to note that the results of ANOVA Type-II and Type-III are the same for such additive models.

Plenty of prior work show that the parameters of classification techniques have an impact on the performance of defect models [27, 48, 58, 59, 83]. While we use a default trees of 100 for random forest models, recent studies [40, 83, 91] show that the parameters of random forest are insensitive to the performance of defect models. Thus, the parameters of random forest models do not pose a threat to validity of our study.

### 7.2 Internal Validity

We studied a limited number of model interpretation techniques. Thus, our results may not generalize to other model interpretation techniques. Nonetheless, other model interpretation techniques can be explored in future work. We provide a detailed methodology for others who would like to re-examine our findings using unexplored model interpretation techniques.

### 7.3 External Validity

The analyzed datasets are part of several corpora (e.g., NASA and PROMISE) of systems that span both proprietary and open source domains. However, we studied a limited number of defect datasets. Thus, the results may not generalize to other datasets and domains. Nonetheless, additional replication studies are needed.

The conclusions of our case study rely on one defect prediction scenario (i.e., within-project defect models). However, there are a variety of defect prediction scenarios in the literature (e.g., cross-project defect prediction [15, 94], just-in-time defect prediction [42], heterogenous defect prediction [67]). Therefore, the practical guidelines may differ for other scenarios. Thus, future research should revisit our study in other scenarios of defect models.

## 8 CONCLUSIONS

In this paper, we set out to investigate (1) the impact of correlated metrics on the interpretation of defect models; (2) the benefits of removing correlated metrics on the interpretation of defect models; and (3) the costs of removing correlated metrics on the performance and stability of defect models. Through a case study of 15 publicly-available defect datasets of systems that span both proprietary and open source domains, we conclude that (1) correlated metrics impact the ranking of metrics according to the 9 studied built-in interpretation techniques for logistic regression and random forest. On the other hand, we find that removing all correlated metrics (2) improves the consistency of the produced rankings regardless of the ordering of metrics (except for ANOVA Type-I); (3) improves the consistency of the highest ranked metric among the studied interpretation techniques; and (4) negligibly impacts the AUC, F-measure, MCC performance, and stability of defect models, suggesting that the benefits of removing correlated metrics outweigh the costs.

Based on our findings, we make the following suggestions for researchers and practitioners:

1) Researchers must mitigate (e.g., remove) correlated metrics prior to constructing a defect model.

2) Researchers must avoid using ANOVA Type-I even if all correlated metrics are removed.

Finally, we would like to emphasize that the goal of this work is not to claim the generalization of our results for every dataset and every analytical model in software engineering. Instead, the key message of our study is to shed light that, for some of the most widely used defect datasets in our field, removing correlated metrics produces more accurate and reliable interpretation of defect models, while leading to a negligible to small impact in the performance and stability of the models. Hence, we recommend that researchers mitigate correlated metrics prior to constructing analytical models, and avoid using ANOVA Type-I (one of the most commonly-used interpretation techniques in software engineering today). Due to the variety of the built-in interpretation techniques and their settings, our paper highlights the essential need for future research to report the exact specification of their models and settings of the used interpretation techniques.